\documentclass[twocolumn,aps,prb,groupedaddress,showpacs]{revtex4}
\usepackage{amsmath}
\usepackage{graphicx}
\usepackage{dcolumn}
\usepackage{bm}

\begin{document}

\title{Two-Fluxon Dynamics in Annular Josephson Junction}
\author{Abdufarrukh A. Abdumalikov, Jr.$^\dag$, Boris A. Malomed$^\ddag$ and
Alexey V. Ustinov$^\dag$}
\date{\today}

\affiliation{$^\dag$Physikalisches Institut III, Universit\"at
Erlangen-N\"urnberg
         Erlangen D-91058, Germany}
\affiliation{$^\ddag$Department of Interdisciplinary Studies,
Faculty of Engineering,
          Tel Aviv University, Tel Aviv 69978, Israel}

\begin{abstract}
Two-fluxon state in an annular Josephson junction in the presence
of external magnetic field is studied analytically, numerically
and experimentally. We obtain an analytical expression for the
potential of interaction between the fluxons moving at arbitrary
velocities (without the use of the ``nonrelativistic"
approximation). Treating the fluxons as quasi-particles, we then
derive equations of motion for them. Direct simulations of the
full extended sine-Gordon equation are in good agreement with
results produced by the analytical model, in a relevant parameter
region. Experimental data qualitatively agree with the numerical
results.
\end{abstract}

\maketitle


\section{\label{intro}Introduction}

A magnetic flux quantum (fluxon) in long Josephson junction (LJJ)
is a well-known physical example of a sine-Gordon soliton.
Ring-shaped (annular) LLJs serve as the ideal setting to study the
fluxon dynamics, as it is not perturbed by boundary conditions at
edges, which is the case for linear LLJs \cite{ring}. Due to the
magnetic-flux quantization in a superconducting ring, the number
of fluxons initially trapped in an annular junction is conserved.
An effective tool, which makes it possible to create an effective
spatially periodic potential for a fluxon trapped in the annular
LJJ, is external dc magnetic field directed parallel to the ring's
plane \cite{Gron}. If $\theta $ is the angular fluxon coordinate
along the ring, the effective potential is $U(\theta )\sim H\cos
\theta $, where $H$ is the strength of the magnetic field. The
minimum of the potential is located at the spot where the fluxon's
magnetic moment is directed along the external field. The dynamics
of a single fluxon in the spatially periodic potential has been an
object of intensive theoretical and experimental investigations
\cite{McLSco,MkrtchyanSmidth,GolubUstin,KivMal89,Ust,MarMon,UstinMalomThyss,CapCos,GSK}.

\begin{figure}[tbp]
\includegraphics{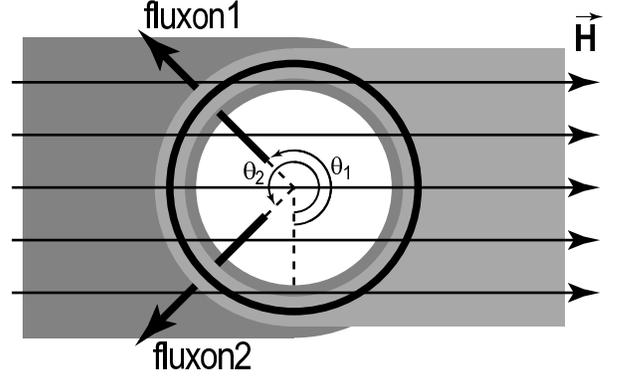}
\caption{The schematic view of an annular Josephson junction with
two trapped fluxons; the magnetic field $\vec{H}$ is applied in
the plane of the junction. Josephson tunnel barrier is shown by
thick black line; in grey are shown superconducting electrodes,
which are extended in the junction plane in order to feed the bias
current.} \label{junction}
\end{figure}

The objective of the present work is to study, both theoretically
and experimentally, dynamics of two fluxons with equal polarities
trapped in the annular LJJ. The system is schematically shown in
Fig.~\ref{junction}. This problem is distinguished by the
interplay of the above-mentioned effective periodic potential
acting on each fluxon and direct interaction (repulsion) between
them.

The interaction between two fluxons in LJJs was first studied
theoretically by Karpman \textit{et al}. (see
Ref.~\onlinecite{KRS} and references therein). These authors found
an analytic expression for the interaction force between fluxons
in the case of a small relative velocity, which corresponds to the
``nonrelativistic'' approximation when the relative velocity is
much smaller than the limit velocity in the LJJ (the Swihart
velocity). However, in the situation relevant to experiment, the
latter condition is not met, in the general case. In this paper,
we aim to develop an analytical description of the interaction
valid in the general (``relativistic'') case. The theory will be
based on an asymptotic method for weakly interacting solitons in
non-integrable systems \cite {GorOst,KivMal89,progress}. The
analysis will be followed by direct simulations and presentation
of experimental results.

The theoretical model for the annular LJJ in the external magnetic
field was proposed in Ref. \cite{Gron}. It is based on the
following extended (perturbed) sine-Gordon (sG) equation for the
superconducting phase difference $\varphi $ between electrodes of
the junction:
\begin{equation}
\varphi _{xx}-\varphi _{tt}-\sin \varphi =\alpha \varphi
_{t}+\gamma +h\sin (qx)\,.  \label{SG}
\end{equation}
Here $x$ is the coordinate along the ring, which is normalized to
the Josephson penetration depth $\lambda _{J}$, the time $t$ is
normalized to the inverse plasma frequency $\omega _{0}^{-1}$,
$\alpha $ is a dissipation coefficient due to the quasiparticle
tunnelling across the junction, and $\gamma $ is the bias current
density (which is necessary in order to compensate losses due to
dissipation), normalized to the critical current density $j_{c}$
of the junction. As usually, the bias current is assumed to be
uniformly distributed along the ring. Further, $q\equiv 2\pi /L$,
where $L$ is the normalized circumference of the junction, and $h$
is the strength of the external magnetic field $H$, normalized by
a sample-specific geometric factor \cite{Gron,MarMon}. If $N$
fluxons are trapped in the ring, Eq.~(\ref{SG}) is supplemented by
the boundary condition
\begin{equation*}
\varphi (L+x,t)=\varphi (x,t)+2\pi N.
\end{equation*}

The paper is organized as following. The single-fluxon dynamics in
the annular LJJ is reviewed in Sec.~\ref{BM}. In
Sec.~\ref{interaction}, the derivation of an effective force of
interaction between two fluxons, valid in the general
(relativistic) case, is presented. Result of numerical
calculations are displayed in Sec.~\ref{numerics}. In
Sec.~\ref{rescondsec}, we analytically consider a special case of
an ostensible resonance in the two-fluxon system. In
Sec.~\ref{experiment}, we present experimental results for two
fluxons trapped in an annular LJJ.

\section{\label{theory}Theory}

\subsection{\label{BM}The basic model}

In our theoretical approach, we assume, as it is usually done,
that the fluxons are well separated from each other, $|\xi
_{1}-\xi _{2}|\gg 1$, where $\xi _{1,2}$ are coordinates of their
centers. In this case the two-soliton state may be represented by
a linear combination of two single-soliton solutions:
\begin{equation}
\varphi =\varphi _{1}+\varphi _{2},  \label{twof}
\end{equation}
where
\begin{equation}
\varphi _{n}=4\arctan \,\exp \left( -\frac{x-\xi
_{n}(t)}{\sqrt{1-\dot{\xi} _{n}^{2}}}\right) +2\pi (n-1)
\label{kink}
\end{equation}
is the single-soliton solution of the unperturbed sG equation, and
$\dot{\xi} _{n}$ is the velocity of the $n$-th soliton. The last
term in Eq. (\ref{kink}) is an arbitrary shift of the background
phase (which is always possible if it is a multiple of $2\pi $),
which is chosen for convenience in what follows below.

Before discussing the interaction of two fluxons, we briefly
recall known results for a single fluxon trapped in an annular
LJJ. This fluxon may be considered as a quasiparticle obeying the
well-known equation of motion \cite {McLSco, Gron91}:

\begin{equation}
\frac{d}{dt}\left( \frac{\dot{\xi}}{\sqrt{1-\dot{\xi}^{2}}}\right)
+\frac{ \alpha \dot{\xi}}{\sqrt{1-\dot{\xi}^{2}}}+\frac{\pi
h}{4}\mathrm{sech}\left( \frac{\pi ^{2}}{L}\right) \sin \left(
q\xi \right) =\frac{\pi \gamma }{4}\,. \label{one}
\end{equation}
It is tantamount to the equation of motion for a relativistic
pendulum in a lossy medium under the action of constant torque.
This equation has solutions of two types. The first type gives
rise to $|\xi (t)|$ growing infinitely. It describes progressive
motion (rotation) of the fluxon around the ring, with a nonzero
mean value of the velocity $\dot{\xi}$. Solutions of the second
type correspond to small oscillations of the fluxon around the
minimum of the effective potential with the frequency
\begin{equation}
\omega _{0}=\left[ \frac{\pi ^{2}}{2L}\mathrm{sech}\left(
\frac{\pi ^{2}} {L} \right) \sqrt{h^{2}-\gamma ^{2}\cosh \left(
\frac{\pi ^{2}}{L}\right)} \right] ^{1/2},  \label{oscilfreq}
\end{equation}
the average velocity being zero. This state exists if $|\gamma |$
is below the critical value,
\begin{equation}
\gamma _{c}^{(1)}=h\,\mathrm{sech}\left( \frac{\pi ^{2}}{L}\right)
. \label{critcurr}
\end{equation}

In the presence of dissipation $(\alpha \not=0)$, the oscillations
are damped, and in the stationary state the fluxon is at rest. On
the other hand, the progressive motion remains possible if the
dissipation is not too strong. The fluxon moves with the average
velocity, which, in the first approximation, is given by the
McLaughlin-Scott formula \cite{McLSco},
\begin{equation}
<\dot{\xi}_{0}>\,=\left( 1+\frac{4\alpha }{\pi \gamma }\right)
^{-1/2}. \label{velocity}
\end{equation}
Equation (\ref{velocity}) determines the current-voltage
characteristics of the junction with a single trapped fluxon.

In the case of two trapped fluxons, we assume that they are
quasiparticles interacting with a certain force (see below), and
all the forces in Eq.~(\ref {one}) act on each fluxon separately.
In this case, three different dynamical regimes are expected: (i)
oscillations of both fluxons (``O-O'' regime, i.e., a zero-voltage
state), (ii) rotation of both fluxons (``R-R'' regime), (iii)
rotation of one fluxon and oscillations of the other one (``R-O''
regime). Note that in R-O regime oscillations take place even in
the presence of dissipation, because the fluxon whose average
velocity is zero is periodically excited by collisions with the
rotating one. All the regimes are implied to be quasi-stationary
states due to the stabilized effect of the dissipation.

\subsection{\label{interaction}The interaction force}

In order to find the interaction force between two fluxons, we use
the center-of-mass reference frame (C-frame). In this frame, the
two solitons move with velocities $\pm u$, where
\begin{equation*}
u=\frac{1-\dot{\xi}_{1}\dot{\xi}_{2}-\sqrt{1-\dot{\xi}_{1}^{2}}\sqrt{1-\dot{
\xi}_{2}^{2}}}{\dot{\xi}_{1}-\dot{\xi}_{2}}\,.
\end{equation*}
In the subsequent calculation, we set $\xi _{1}<\xi _{2}$.

To calculate an effective potential of the interaction between
fluxons, which will then produce the interaction force, we start
with the Hamiltonian of the unperturbed sG equation for the
infinitely long system:
\begin{equation}
H=\int\limits_{-\infty }^{\infty }dx\left( \frac{1}{2}\varphi
_{t}^{2}+\frac{ 1}{2}\varphi _{x}^{2}+1-\cos \varphi \right) .
\label{H}
\end{equation}
For the calculation of the Hamiltonian (\ref{H}), we divide the
space into two parts. The left (right) part occupies the space
from $-\infty$ ($+\infty $) to the midpoint between the two
fluxons, $a\equiv (\xi _{1}+\xi _{2})/2$. We perform the actual
calculation for the left part only, as for the right part the
calculation is quite similar.

In the left part, we substitute the solution as the linear
combination (\ref {twof}), where $\varphi _{2}$ is considered as a
small perturbation, as the two fluxons are assumed to be well
separated. Then, the Hamiltonian is written, in the first
approximation, as
\begin{equation*}
H_{\mathrm{left}}=H_{1}+\delta H_{\mathrm{left,int}}\,,
\end{equation*}
where $H_{1}=8/\sqrt{1-u^{2}}$ is the Hamiltonian of the
unperturbed sG soliton, and the interaction term is of the first
order with respect to the weak field $\varphi _{2}\,$,
\begin{equation}
\delta H_{\mathrm{left,int}}=\int\limits_{-\infty }^{a}dx\left(
\varphi _{1,x}\varphi _{2,x}+\varphi _{1,t}\varphi _{2,t}+\varphi
_{2}\sin \varphi _{1}\right) \,  \label{hint}
\end{equation}
(recall the subscripts $x$ and $t$ stand for the corresponding
partial derivatives). Substituting the expression valid for a
moving soliton, $\varphi _{n,t}=-\dot{\xi}\varphi _{n,x}$, and
integrating by parts, we obtain

\begin{multline}
\delta H_{\mathrm{left,int}}=(1-u^{2})\varphi _{1,x}\varphi
_{2}|_{-\infty }^{a}+ \\ +\int\limits_{-\infty }^{a}dx\varphi
_{2}\left( -\varphi _{1,xx}+\varphi _{1,tt}+\sin \varphi
_{1}\right) \,.  \label{integral}
\end{multline}
The integral term in Eq. (\ref{integral}) is zero as the bracketed
expression is the sine-Gordon equation proper (this way to nullify
the integral terms is known in the general analysis of the
interaction between separated solitons \cite{progress}). The
contribution to the first term in Eq. (\ref{integral}) at the left
limit, $x=-\infty $, is zero too because both $\varphi _{1,x}$ and
$\varphi _{2}$ decay exponentially at infinity. In order to
calculate the contribution from the upper limit, we use the
asymptotic form of $\varphi _{1,x}$ ($\varphi _{2}$) at large
values of $x$ (which is also a known point in the general analysis
of the soliton-soliton interaction \cite{progress}):

\begin{gather}
\varphi _{1,x}=-\frac{4}{\sqrt{1-u^{2}}}\exp \left( -\frac{x-\xi
_{1}}{\sqrt{
1-u^{2}}}\right) ,  \notag \\
\varphi _{2}=-4\exp \left( \frac{x-\xi
_{2}}{\sqrt{1-u^{2}}}\right) \,. \label{asympt}
\end{gather}

After the substitution of the expressions (\ref{asympt}) into Eq.
(\ref {integral}) and calculation of the nonvanishing contribution
to the first term from the right limit, $x=a$, and then getting
back from the C-frame to the laboratory reference frame (L-frame),
we arrive at an expression for the interaction potential for two
moving fluxons in the infinitely long junction,
\begin{equation}
\delta
H_{\mathrm{int}}=32\frac{\sqrt{1-u^{2}}}{\sqrt{1-V^{2}}}\exp
\left(- \frac{|\xi _{1}-\xi _{2}|}{\sqrt{1-u^{2}}}\right) ,
\label{potential}
\end{equation}
where the contribution of the right half space is taken into
account and
\begin{equation}
V=\frac{1+\dot{\xi}_{1}\dot{\xi}_{2}-\sqrt{1-\dot{\xi}_{1}^{2}}\sqrt{1-\dot{
\xi}_{2}^{2}}}{\dot{\xi}_{1}+\dot{\xi}_{2}}
\end{equation}
is the velocity of the center-of-mass of the two-fluxon set in
L-frame. In the case of equal velocities, the potential
(\ref{potential}) reduces to the well-known result of Karpman
\textit{et al}. \cite{KRS}.

The potential (\ref{potential}) gives rise to two forces acting on
each soliton, due to the ring geometry of the system, which should
be added to the individual forces in Eq. (\ref{one}),
\begin{multline}\label{force}
\left( F_{\mathrm{int}}\right) _{1}=-\left(
F_{\mathrm{int}}\right) _{2}=- \frac{1}{8}\frac{d}{d\Delta
x}\delta H_{\mathrm{int}}^{\mathrm{ring}}=
\\
\frac{4}{\sqrt{1-V^{2}}}\left[ \exp \left( -\frac{\Delta
X}{1-u^{2}}\right) -\exp \left( -\frac{L-\Delta X}{1-u^{2}}\right)
\right],
\end{multline}
where $\Delta X=|\xi _{1}-\xi _{2}|\,$\texttt{,} and $8$ is for
the effective mass of the fluxon in the present notation.
Equations (\ref{one}) and (\ref{force}) describe the dynamics of
the two-fluxon system in an annular LJJ. These equations of motion
we solved numerically by means of the fourth-order Runge-Kutta
method.

\section{\label{numerics}Numerical calculations}

In order to verify the theory presented above, we checked
numerical solutions of the quasi-particle equations of motion
against direct simulations of the full equation~(\ref{SG}). Here
we present results for fixed values $\alpha =0.02$ and $L=20$,
while the bias current $\gamma $ was varied in steps of $0.002$.

The general behavior of the system can be described as the
following. While $\gamma $ increases form zero, both fluxons
originally stay pinned in the effective potential induced by the
magnetic field, so that the voltage across the junction is zero.
At a critical value of the current $\gamma_{c}$, the system
switches to the R-O regime, in which one of the fluxons rotates,
while the other one oscillates due to periodic collisions with the
moving fluxon. This state is stable up to another critical point,
$\gamma <\gamma _{s}$. On the other hand, decreasing the bias
current leads to a transition to the regime with both fluxons
pinned (zero voltage) at a different value, $\gamma =\gamma _{r}$.

At $\gamma >\gamma _{s}\,$, the system operates in R-R regime with
both fluxons rotating. Further increase of the bias current does
not change the state of the system, up to a large value of the
current, at which the junction switches to the ``whirling'' (alias
resistive) state, with uniform rotation of the phase in all the
system. When decreasing the bias current, the system switches
first to the R-O regime, and then to the zero-voltage state.
\begin{figure}[b]
\includegraphics[scale=0.8]{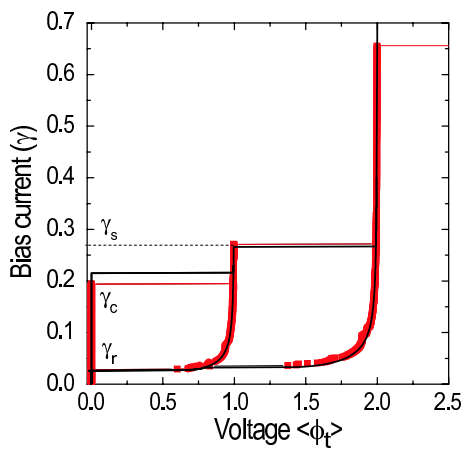}
\caption{The current-voltage characteristics of the long annular
junction with two trapped fluxons found from direct numerical
simulation of Eq.~(\ref {SG}) (dots), and from the analytical
model based on Eq.~(\ref{one}) and (\ref{force})(solid line). In
this case, the magnetic field is fixed to $h=0.3$.} \label{ivc}
\end{figure}

Typical current-voltage (I-V) characteristics, which display all
these states and transitions, are presented in Fig.~\ref{ivc}.
Points shown by dots correspond to the numerical solution of the
full equation~(\ref{SG}), while the lines depict solutions of the
quasi-particle model based on Eqs.~(\ref{one}) and (\ref{force}).
As is seen, the analytical quasi-particle model making use of the
expression (\ref{force}) for the interaction forces is in good
agreement with direct simulations.

\begin{figure}[htb]
\includegraphics[scale=0.8]{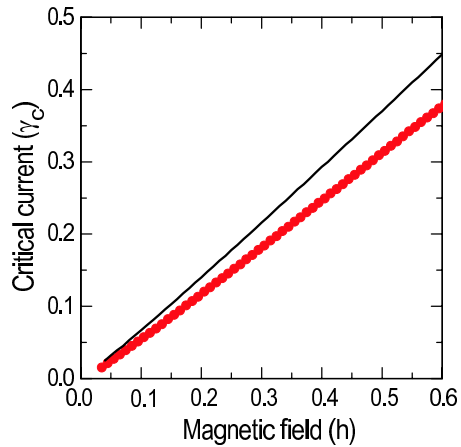}
\caption{The critical current $\protect\gamma _{c}$ versus the
magnetic field $h$, obtained from direct numerical simulations of
the full equation (\ref{SG}) (dots) and from the quasi-particle
model (line).} \label{ich}
\end{figure}

The comparison of the critical values $\gamma _{c}$ of the bias
current, obtained from the full simulations and from the
analytical model, is shown in Fig. \ref{ich}. A small difference
between them can be attributed to the fact that, in the O-O
regime, the actual distance between the pinned fluxons is small,
hence the assumption of far separated fluxons does not apply in
this case. Indeed the numerical simulations show that the distance
between the fluxons in static case changes, depending on the
magnetic field, in range of $0.8-1.5$.

For small currents, the I-V curve of the R-O and R-R regimes,
found from the direct simulations of Eq.~(\ref{SG}), feature
additional small steps, which are due to resonant generation of
radiation by the fluxons moving in the periodic potential
\cite{Ust}.

\begin{figure}[hbt]
\includegraphics[scale=0.8]{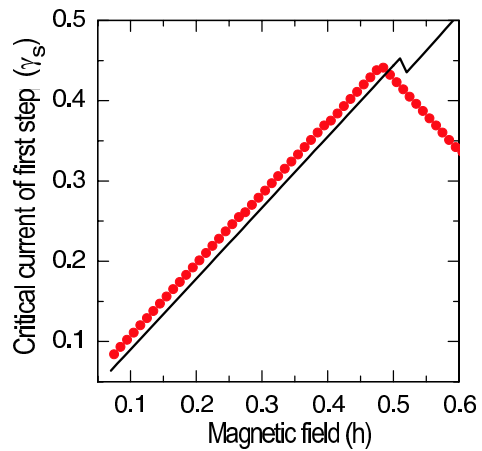}
\caption{The dependence of the maximum current of the first step
in the I-V curve, $\protect\gamma _{s}$, on the magnetic field
$h$, as found from direct simulations of the full equation
(\ref{SG}) (dots), and from the analytical model based on Eqs.
(\ref{one}) and (\ref{force}) (line).} \label{ish}
\end{figure}

The comparison of the other critical value of the bias current,
$\gamma _{s}$ (which corresponds to the first step of the I-V
characteristics), again\ as found from the direct simulations and
analytical model, is shown, versus the magnetic field, in
Fig.~\ref{ish}. At small values of the magnetic field, these
dependencies agree very well. However, for $h>0.45$ the curve
generated by the direct simulations goes down with the increase of
the field. Actually, in this region the parameters $\gamma $ and
$h$ are too large to apply the perturbation theory. With the
further increase of $h$, $\gamma _{s} $ decreases until it becomes
equal to $\gamma _{c}$. For a still stronger field, the system
switches from the O-O regime directly to the whirling state.

On the other hand, the curve produced by the analytical
approximation continues to go up with the field until $h=0.5$. For
fields larger than $0.5$, the system resonantly switches from the
R-O regime to the R-R one, but at so large values of the field the
quasi-particle model based on the perturbation theory becomes
completely irrelevant.

\subsection{\label{rescondsec}A resonance condition}

An noteworthy feature of the two-fluxon dynamics in the R-O regime
is a possibility of a resonance between the natural frequency of
oscillations of the trapped fluxon, and periodic excitation due to
its collisions with the rotating one. The resonance condition is
obtained by equating the frequency of small oscillations $\omega
_{0}$, given by Eq.~(\ref{oscilfreq}), and the rotation frequency
$\omega _{r}=q<\dot{\xi}_{r}>$, where $\dot{\xi}_{r}$ is the
velocity of the rotating fluxon, which can be obtained from Eq.
(\ref {velocity}). This yields

\begin{equation}
h=\gamma _{\mathrm{res}}\cosh \left( \frac{\pi ^{2}}{L}\right)
\sqrt{1+\frac{ 64\pi ^{2}}{L^{2}(\pi \gamma
_{\mathrm{res}}+4\alpha )^{2}}}\,, \label{rescondeq}
\end{equation}
where $\gamma _{\mathrm{res}}$ is the value of the bias current
corresponding to the resonance.

\begin{figure}[hbt]
\includegraphics[scale=0.8]{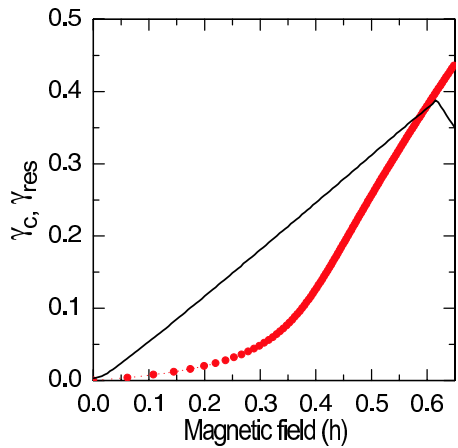}
\caption{Critical current of two fluxon state found from numerical
solutions of Eq.~(\ref{SG}) (solid line) and the current of
expected resonance (dots)} \label{rescond:pic}
\end{figure}

This dependence for $L=20$ and $\alpha=0.02$ is plotted in
Fig.~\ref{rescond:pic} by dots.
The line corresponds to the critical current of the two fluxon
state found from full numerical simulations of Eq.~(\ref{SG}) for
the same length. It may be expected that this resonance would
result in a resonant switching from R-O to R-R regime and drop of
the critical current of the first step $\gamma _s$ vs. magnetic
field $h$. However, at low magnetic fields the system switches to
the R-O branch at higher current than the current corresponding to
the resonance. At large magnetic field it switches directly to R-R
branch, because the total perturbation (field + bias current) is
too strong. Here the perturbation approach is not any more
applicable; the numerical solution of the analytical model
reflects the resonance behavior by a drop on $\gamma_s(h)$
dependence (solid line in Fig.~\ref{ish}). With increase of the
junction length $L$ the the intersection point of the curves
$\gamma_c(h)$ and $\gamma_{\mathrm{res}}(h)$ moves upward in bias
current, while its dependence on magnetic field is very weak.

\section{\label{experiment}Experiment}

\begin{figure}[t]
\includegraphics[scale=0.8]{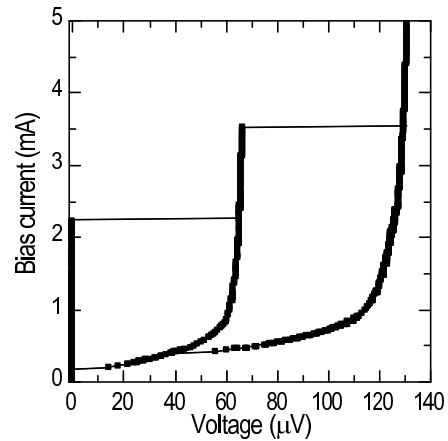}
\caption{Current-voltage characteristics of the annular Josephson
junction with two trapped fluxons.} \label{ivcexp}
\end{figure}

Measurements of the I-V characteristics of the two-fluxon state
were performed in long annular Nb-Al-AlO$_{x}$-Nb junctions. Due
to the magnetic flux quantization in a superconducting ring, the
number of initially trapped fluxons (solitons) is conserved.
Trapping of a magnetic flux in the junction was achieved while
cooling the sample below the critical temperature $T_{c}^{
\mathrm{Nb}}=9.2$~K of niobium, in the presence of a small bias
current passing through the junction. The number of trapped
fluxons was determined from the highes voltage of the highest
resonant branche on the current-voltage characteristics
(Fig.~\ref{ivcexp}). Experiments were performed by applying the
bias current $I$ from top to the bottom electrode of the junction
and measuring the dc voltage generated due to the motion
(\textit{rotation}, in terms of the theoretical consideration) of
the trapped fluxons. The results presented below were obtained for
a junction with the mean diameter $100~\mu \mathrm{m}$ and ring's
width $3~\mu \mathrm{m}$. The circumference (length of the annular
junction) in the normalized units was $L=28.5$, and the effective
loss parameter was estimated as $\alpha =0.03$. The measurements
were performed at 4.2 K.

At zero magnetic field, depinning of a single-fluxon was observed,
as a switching from the zero voltage state to the single-fluxon
step of the I-V curve (in the state with one trapped fluxon), at
the current $I_{c}$ that was smaller by a factor $\approx 65$ than
the critical current for the same junction, measured without
trapped fluxons. This fact indicates a high degree of homogeneity
of the junction (a strong local inhomogeneity would give rise to a
much larger value of the fluxon-depinning critical current).

We measured I-V curves of the state with two trapped fluxons for
different strengths of the applied magnetic field. One of these
curves is shown in Fig.~\ref{ivcexp}. The experimental curves
qualitatively agree with our analysis illustrated by
Fig.~\ref{ivc}. Two branches of the I-V curve are observed. The
first branch, at around $65\ \mu$V corresponds to the R-O regime
and the second one at about $130\ \mu$V to R-R regime.

\begin{figure}[t]
\includegraphics[scale=0.8]{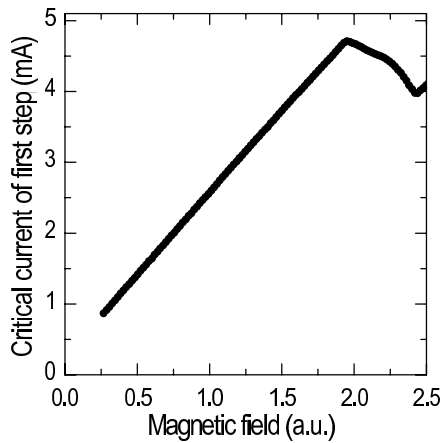}
\caption{The experimentally found critical current for the jump
from the one-fluxon step of the I-V curve to the two-fluxon one,
as a function of the magnetic field.} \label{ishexp}
\end{figure}

Figure~\ref{ishexp} shows the measured value of the switching
current for the transition from the R-O regime to the R-R one,
versus the external magnetic field. The curve is quite similar to
the calculated dependence $\gamma _{s}(h)$ (dots in
Fig.~\ref{ish}), which was obtained above from direct integration
of the full sine-Gordon model (\ref{SG}).

\section{\label{summary}Summary}

We have reported results of theoretical and experimental studies
of two-fluxon dynamics in an long annular Josephson junction in
presence of the external magnetic field. An analytical expression
for the interaction force between two fluxons moving at different
arbitrary velocities has been derived. Solution of the system of
two coupled quasi-particle equations of motion for the fluxons,
taking into account the interaction force, demonstrates good
agreement with direct numerical simulations of the two-fluxon
state in the full sine-Gordon model including all the perturbation
factors.

Three distinct dynamical regimes of the two-fluxon state have been
thus identified. First, both fluxons may be pinned in the
potential induced by the magnetic field. There is some discrepancy
in the prediction of the critical current, which destroys this
static regime, between the quasi-particle model and direct
simulations, due to the fact that the separation between the two
trapped fluxons is rather small, while the analytical model
assumes them to be well separated.

In the second regime, one of the fluxons rotates around the
junction, while the other one oscillates in the potential well.
For this (R-O) regime, the maximum current is very accurately
predicted by the quasi-particle model, if compared to direct
numerical results. For this case, we have also investigated the
possibility of the resonant excitation of the trapped fluxon by
periodic collisions with the rotating one.

In the third regime, both fluxons rotate. The corresponding I-V
curves found by direct simulations demonstrate several additional
small steps, which are due to resonant generation of
small-amplitude plasma waves by fluxons moving in the periodic
potential, which has been analyzed earlier \cite{Ust} for the
single-fluxon case.

The method developed in this work to calculate, in the general
case, the effective interaction force acting between two moving
solitons, may find application to other soliton-bearing systems.

\begin{acknowledgments}
A.A.A. is grateful to F.Kh. Abdullaev and E.N. Tsoy for usefull
discussions. We also wish to thank A. Kemp for help during
experiment. B.A.M. appreciated hospitality of Physikalisches
Institut III, of Universit\"at Erlangen-N\"urnberg.
\end{acknowledgments}


\end{document}